# The stability of the QED vacuum in the temporal gauge

by

Dan Solomon

Rauland-Borg Corporation
3450 W. Oakton
Skokie, IL USA

Email: dan.solomon@rauland.com

**Abstract**

The stability of the vacuum for QED in the temporal gauge will be examined. It is generally assumed that the vacuum state is the quantum state with the lowest energy. However, it will be shown that this is not the case for a system consisting of a fermion field coupled to a quantized electromagnetic field in the temporal gauge. It will be shown that for this situation there exist quantum states with less energy than the vacuum state.

## I. Introduction.

In this article the stability of the QED vacuum in the temporal gauge will be examined. In the temporal gauge the gauge condition is given by the relationship $A_0 = 0$ [1,2,3,4] where $A_0$ is the scalar component of the electric potential. The advantage of the temporal gauge is due to the simplicity of the commutation relationship between the electromagnetic field quantities which are given below. In the coulomb gauge, for instance, these are more complicated. Due to this fact the temporal gauge is particularly useful in the treatments of QED which use the functional Schrödinger equation [3,4].

We will examine a field consisting of fermions coupled to a quantized electromagnetic field. It is generally assumed that that there is a lower bound to the energy of quantum states and that the quantum state with the lowest energy is the vacuum state. We will show that that this is not the case for QED in the temporal gauge and that there must exist quantum states with less energy than the vacuum state. Throughout this discussion we will use $\hbar = c = 1$. Also vectors are indicated by bold text.

We will work in the Schrödinger picture. In this case the field operators are time independent and the time dependence of the quantum system is reflected in the state vector $|\Omega(t)\rangle$ which evolves in time according to the Schrödinger equation,

$$i\frac{\partial|\Omega(t)\rangle}{\partial t} = \hat{H}|\Omega(t)\rangle \tag{1.1}$$

where the Hamiltonian $\hat{H}$ is [3],

$$\hat{H} = \hat{H}_{0,D} + \hat{H}_{0,M} - \int \hat{\mathbf{J}}(\mathbf{x})\cdot\hat{\mathbf{A}}(\mathbf{x})d\mathbf{x} \tag{1.2}$$

The quantities in the above expression are defined by,

$$\hat{H}_{0,D} = \frac{1}{2}\int\left[\hat{\psi}^{\dagger}(\mathbf{x}), H_{0,D}\hat{\psi}(\mathbf{x})\right]d\mathbf{x}; \quad H_{0,D} = -i\boldsymbol{\alpha}\cdot\nabla + \beta m \tag{1.3}$$

$$\hat{H}_{0,M} = \frac{1}{2}\int\left(\hat{\mathbf{E}}^2 + \hat{\mathbf{B}}^2\right)d\mathbf{x}; \qquad \hat{\mathbf{B}}(\mathbf{x}) = \nabla\times\hat{\mathbf{A}}(\mathbf{x}) \tag{1.4}$$

$$\hat{\mathbf{J}}(\mathbf{x}) = \frac{q}{2}\left[\hat{\psi}^{\dagger}(\mathbf{x}), \boldsymbol{\alpha}\hat{\psi}(\mathbf{x})\right] \tag{1.5}$$

In the above expressions m is the fermion mass, $\boldsymbol{\alpha}$ and $\beta$ are the usual 4x4 matrices, q is the electric charge, $\hat{H}_{0,D}$ is the Dirac Hamiltonian, $\hat{H}_{0,M}$ is the Hamiltonian for the electromagnetic field, and $\hat{\mathbf{J}}(\mathbf{x})$ is the current operator. The Schrödinger picture time independent fermion field operators are $\hat{\psi}(\mathbf{x})$ and $\hat{\psi}^{\dagger}(\mathbf{x})$. The field operators for the electromagnetic field are $\hat{\mathbf{A}}(\mathbf{x})$ and $\hat{\mathbf{E}}(\mathbf{x})$. The electromagnetic field operators are real so that $\hat{\mathbf{A}}^{\dagger}(\mathbf{x}) = \hat{\mathbf{A}}(\mathbf{x})$ and $\hat{\mathbf{E}}^{\dagger}(\mathbf{x}) = \hat{\mathbf{E}}(\mathbf{x})$.

The field operators obey the following relationships [2,3],

$$\left[\hat{A}^i(\mathbf{x}), \hat{E}^j(\mathbf{y})\right] = -i\delta_{ij}\delta^3(\mathbf{x}-\mathbf{y}); \; \left[\hat{A}^i(\mathbf{x}), \hat{A}^j(\mathbf{y})\right] = \left[\hat{E}^i(\mathbf{x}), \hat{E}^j(\mathbf{y})\right] = 0 \tag{1.6}$$

and

$$\left\{\hat{\psi}_a^{\dagger}(\mathbf{x}), \hat{\psi}_b(\mathbf{y})\right\} = \delta_{ab}\delta(\mathbf{x}-\mathbf{y}); \; \left\{\hat{\psi}_a^{\dagger}(\mathbf{x}), \hat{\psi}_b^{\dagger}(\mathbf{y})\right\} = \left\{\hat{\psi}_a(\mathbf{x}), \hat{\psi}_b(\mathbf{y})\right\} = 0 \tag{1.7}$$

where "a" and "b" are spinor indices. In addition, all commutators between the electromagnetic field operators and fermion field operators are zero, i.e.,

$$\left[\hat{\mathbf{A}}(\mathbf{x}), \hat{\psi}(\mathbf{y})\right] = \left[\hat{\mathbf{E}}(\mathbf{x}), \hat{\psi}(\mathbf{y})\right] = \left[\hat{\mathbf{A}}(\mathbf{x}), \hat{\psi}^{\dagger}(\mathbf{y})\right] = \left[\hat{\mathbf{E}}(\mathbf{x}), \hat{\psi}^{\dagger}(\mathbf{y})\right] = 0 \tag{1.8}$$

Define,

$$\hat{G}(\mathbf{x}) = \nabla\cdot\hat{\mathbf{E}}(\mathbf{x}) - \hat{\rho}(\mathbf{x}) \tag{1.9}$$

where the current operator $\hat{\rho}(\mathbf{x})$ is defined by,

$$\hat{\rho}(\mathbf{x}) = \frac{q}{2}\left[\hat{\psi}^{\dagger}(\mathbf{x}), \hat{\psi}(\mathbf{x})\right] \tag{1.10}$$

The state vector $|\Omega(t)\rangle$ must satisfy the gauss's law constraint [3],

$$\hat{G}(\mathbf{x})|\Omega(t)\rangle = 0 \tag{1.11}$$

It can be shown [1] that $\left[\hat{H}, \hat{G}(\mathbf{x})\right] = 0$ therefore if (1.11) is valid at some initial time it will be valid for all time.

The energy of a normalized state $|\Omega\rangle$ is given by,

$$E(|\Omega\rangle) = \langle\Omega|\hat{H}|\Omega\rangle \tag{1.12}$$

where the normalization condition is $\langle\Omega|\Omega\rangle = 1$. The commonly held assumption is that there exists a state $|\Omega_{vac}\rangle$, called the vacuum state, which is a state of minimum energy so that all other states have an energy that is greater than the vacuum state, i.e.,

$$E(|\Omega\rangle) - E(|\Omega_{vac}\rangle) > 0 \text{ for all } |\Omega\rangle \neq |\Omega_{vac}\rangle \tag{1.13}$$

We will show that the above statement cannot be true if there exists a normalized state $|\Omega_1\rangle$ which satisfies Gauss's law and for which the divergence of the current expectation value is non-zero, that is,

$$\nabla \cdot \langle\Omega_1|\mathbf{J}(\mathbf{x})|\Omega_1\rangle \neq 0 \text{ in some region of space.} \tag{1.14}$$

Before proceeding we must ask the question "how do we know that a state $|\Omega_1\rangle$ can be found where the above condition holds?". The answer is that if quantum mechanics is a correct model of the real world then there must exist many states where the above condition holds because in the real world there are many examples where the divergence of the current is non-zero over some region of space.

Now given some initial state new states can be defined by acting with operators on the initial state [6]. With this in mind define the state,

$$|\Omega_2\rangle = e^{-i\hat{C}}|\Omega_1\rangle \tag{1.15}$$

where the operator $\hat{C}$ is defined by,

$$\hat{C} = \int \hat{\mathbf{E}}(\mathbf{x}) \cdot \nabla\chi(\mathbf{x}) d\mathbf{x} \tag{1.16}$$

and where $\chi(\mathbf{x})$ is an arbitrary real valued function. Note that dual state is,

$$\langle\Omega_2| = \langle\Omega_1|e^{+i\hat{C}^\dagger} = \langle\Omega_1|e^{+i\hat{C}} \tag{1.17}$$

where we have used $\hat{C}^\dagger = \hat{C}$ since $\hat{\mathbf{E}}(\mathbf{x})$ and $\chi(\mathbf{x})$ are both real. From this we have that $\langle\Omega_2|\Omega_2\rangle = \langle\Omega_1|\Omega_1\rangle = 1$ where we use the relationship,

$$e^{+i\hat{C}}e^{-i\hat{C}} = 1 \tag{1.18}$$

Now is $|\Omega_2\rangle$ a valid state, i.e., does it satisfy (1.11)? Based on the commutator relationships (1.6) and (1.8) we see that the operator $\hat{C}$ commutes with both $\hat{\mathbf{E}}(\mathbf{x})$ and $\hat{\rho}(\mathbf{x})$. Therefore $\hat{G}(\mathbf{x})|\Omega_2\rangle = e^{-i\hat{C}}\hat{G}(\mathbf{x})|\Omega_1\rangle = 0$ so that $|\Omega_2\rangle$ satisfies (1.11) since $|\Omega_1\rangle$ has been assumed to satisfy $\hat{G}(\mathbf{x})|\Omega_1\rangle = 0$.

Next we want to evaluate the energy of the state $|\Omega_2\rangle$. To do this use (1.2) and (1.12) to obtain,

$$E(|\Omega_2\rangle) = \langle\Omega_2|\hat{H}_{0,D}|\Omega_2\rangle + \langle\Omega_2|\hat{H}_{0,M}|\Omega_2\rangle - \langle\Omega_2|\int \hat{\mathbf{J}}(\mathbf{x}) \cdot \hat{\mathbf{A}}(\mathbf{x}) d\mathbf{x}|\Omega_2\rangle \tag{1.19}$$

Consider first the term $\langle\Omega_2|\hat{H}_{0,D}|\Omega_2\rangle$. To evaluate this use the fact that $\hat{\mathbf{E}}(\mathbf{x})$, and thereby $\hat{C}$, commutes with the fermion field operators $\hat{\psi}(\mathbf{x})$ and $\hat{\psi}^{\dagger}(\mathbf{x})$. Use this fact along with (1.18) to obtain,

$$\langle\Omega_2|\hat{H}_{0,D}|\Omega_2\rangle = \langle\Omega_1|\hat{H}_{0,D}|\Omega_1\rangle \tag{1.20}$$

Next consider the term $\langle\Omega_2|\hat{H}_{0,M}|\Omega_2\rangle$. From (1.6) we obtain,

$$\left[\hat{\mathbf{A}}(\mathbf{x}),\hat{C}\right] = -i\nabla\chi(\mathbf{x}) \tag{1.21}$$

Use this result to obtain,

$$\left[\hat{\mathbf{B}}(\mathbf{x}),\hat{C}\right] = \nabla\times\left[\hat{\mathbf{A}}(\mathbf{x}),\hat{C}\right] = -i\nabla\times\nabla\chi(\mathbf{x}) = 0 \tag{1.22}$$

Therefore $\hat{C}$ commutes with $\hat{H}_{0,M}$ so that,

$$\langle\Omega_2|\hat{H}_{0,M}|\Omega_2\rangle = \langle\Omega_1|\hat{H}_{0,M}|\Omega_1\rangle \tag{1.23}$$

Now for last term in (1.19) use the fact that $\hat{C}$ commutes with $\hat{\mathbf{J}}(\mathbf{x})$ to obtain,

$$\langle\Omega_2|\int\hat{\mathbf{J}}(\mathbf{x})\cdot\hat{\mathbf{A}}(\mathbf{x})d\mathbf{x}|\Omega_2\rangle = \langle\Omega_1|\int\hat{\mathbf{J}}(\mathbf{x})\cdot\left(e^{+i\hat{C}}\hat{\mathbf{A}}(\mathbf{x})e^{-i\hat{C}}\right)d\mathbf{x}|\Omega_1\rangle \tag{1.24}$$

To evaluate the above expression further use the Baker-Campell-Hausdorff relationships [7] which states that,

$$e^{+\hat{O}_1}\hat{O}_2 e^{-\hat{O}_1} = \hat{O}_2 + \left[\hat{O}_1,\hat{O}_2\right] + \frac{1}{2}\left[\hat{O}_1,\left[\hat{O}_1,\hat{O}_2\right]\right] + \dots \tag{1.25}$$

where $\hat{O}_1$ and $\hat{O}_2$ are operators. Use this relationship along with (1.6) and (1.21) to obtain,

$$e^{+i\hat{C}}\hat{\mathbf{A}}(\mathbf{x})e^{-i\hat{C}} = \hat{\mathbf{A}}(\mathbf{x}) - \nabla\chi(\mathbf{x}) \tag{1.26}$$

Use this result in (1.24) to obtain,

$$\langle\Omega_2|\int\hat{\mathbf{J}}(\mathbf{x})\cdot\hat{\mathbf{A}}(\mathbf{x})d\mathbf{x}|\Omega_2\rangle=\langle\Omega_1|\int\hat{\mathbf{J}}(\mathbf{x})\cdot\left(\hat{\mathbf{A}}(\mathbf{x})-\nabla\chi(\mathbf{x})\right)d\mathbf{x}|\Omega_1\rangle \tag{1.27}$$

Use the above results in (1.19) to yield,

$$\mathrm{E}\left(|\Omega_2\rangle\right)=\langle\Omega_1|\left(\hat{\mathrm{H}}_{0,\mathrm{D}}+\hat{\mathrm{H}}_{0,\mathrm{M}}-\int\hat{\mathbf{J}}(\mathbf{x})\cdot\hat{\mathbf{A}}(\mathbf{x})d\mathbf{x}\right)|\Omega_1\rangle+\int\langle\Omega_1|\hat{\mathbf{J}}(\mathbf{x})|\Omega_1\rangle\cdot\nabla\chi(\mathbf{x})d\mathbf{x} \tag{1.28}$$

Next use (1.2) and (1.12) in the above and integrate the last term by parts, assuming reasonable boundary conditions, to obtain,

$$\mathrm{E}\left(|\Omega_2\rangle\right)=\mathrm{E}\left(|\Omega_1\rangle\right)-\int\chi(\mathbf{x})\nabla\cdot\langle\Omega_1|\hat{\mathbf{J}}(\mathbf{x})|\Omega_1\rangle d\mathbf{x} \tag{1.29}$$

Next subtract the energy of the vacuum state, $\mathrm{E}\left(|\Omega_{\mathrm{vac}}\rangle\right)$, from both sides to obtain,

$$\mathrm{E}\left(|\Omega_2\rangle\right)-\mathrm{E}\left(|\Omega_{\mathrm{vac}}\rangle\right)=\left(\mathrm{E}\left(|\Omega_1\rangle\right)-\mathrm{E}\left(|\Omega_{\mathrm{vac}}\rangle\right)\right)-\int\chi(\mathbf{x})\nabla\cdot\langle\Omega_1|\hat{\mathbf{J}}(\mathbf{x})|\Omega_1\rangle d\mathbf{x} \tag{1.30}$$

Now in the above expression the quantities $\left(\mathrm{E}\left(|\Omega_1\rangle\right)-\mathrm{E}\left(|\Omega_{\mathrm{vac}}\rangle\right)\right)$ and $\langle\Omega_1|\hat{\mathbf{J}}(\mathbf{x})|\Omega_1\rangle$ are independent of $\chi(\mathbf{x})$. Recall that we have picked the quantum state $|\Omega_1\rangle$ so that $\nabla\cdot\langle\Omega_1|\hat{\mathbf{J}}(\mathbf{x})|\Omega_1\rangle$ is nonzero. Based on this we can always find a $\chi(\mathbf{x})$ so that $\left(\mathrm{E}\left(|\Omega_2\rangle\right)-\mathrm{E}\left(|\Omega_{\mathrm{vac}}\rangle\right)\right)$ is a negative number. For example, let $\chi(\mathbf{x})=\lambda\nabla\cdot\langle\Omega_1|\hat{\mathbf{J}}(\mathbf{x})|\Omega_1\rangle$ where $\lambda$ is a constant. Then (1.30) becomes,

$$\mathrm{E}\left(|\Omega_2\rangle\right)-\mathrm{E}\left(|\Omega_{\mathrm{vac}}\rangle\right)=\left(\mathrm{E}\left(|\Omega_1\rangle\right)-\mathrm{E}\left(|\Omega_{\mathrm{vac}}\rangle\right)\right)-\lambda\int\left(\nabla\cdot\langle\Omega_1|\hat{\mathbf{J}}(\mathbf{x})|\Omega_1\rangle\right)^2 d\mathbf{x} \tag{1.31}$$

Now, since $\nabla\cdot\langle\Omega_1|\hat{\mathbf{J}}(\mathbf{x})|\Omega_1\rangle$ is nonzero, the integral must be positive so that as $\lambda\rightarrow\infty$ the quantity $\left(\mathrm{E}\left(|\Omega_2\rangle\right)-\mathrm{E}\left(|\Omega_{\mathrm{vac}}\rangle\right)\right)\rightarrow-\infty$. Therefore the energy of the state $|\Omega_2\rangle$ is less than that of the vacuum state $|\Omega_{\mathrm{vac}}\rangle$ by an arbitrarily large amount. Therefore there is no lower bound to the energy of a QED quantum state in the temporal gauge.

## II. Interaction with Classical fields

In the previous section we have shown that if there exists a state $|\Omega_1\rangle$ that satisfies (1.14) then there exists a state $|\Omega_2\rangle$ whose energy is less that that of $|\Omega_1\rangle$ by an arbitrarily large amount. This suggests the possibility that it would be possible to extract an arbitrarily large amount of energy from a quantum state through the interaction with an external field. It will be shown in this section that this is theoretically possible.

In the absence of external interactions the energy of a quantum state remains constant. In order to change the energy we must allow the field operators to interact with external sources or fields. This is done by adding an interaction term to the Hamiltonian. Let this term be,

$$\hat{H}_{int} = -\int \mathbf{S}(\mathbf{x},t)\cdot\hat{\mathbf{A}}(\mathbf{x})d\mathbf{x} - \int \hat{\mathbf{J}}(\mathbf{x})\cdot\mathbf{R}(\mathbf{x},t)d\mathbf{x} \tag{2.1}$$

In the above expression $\mathbf{S}(\mathbf{x},t)$ is a classical field that interacts with the quantized electromagnetic field and $\mathbf{R}(\mathbf{x},t)$ is a separate classical field that interacts with the fermion current operator. It should not be assumed that the classical fields $\mathbf{S}(\mathbf{x},t)$ and $\mathbf{R}(\mathbf{x},t)$ correspond to physical fields that actually exist. For the purposes of this discussion these fields are fictitious. They have been introduced for the purposes of perturbing the Hamiltonian in order to change the energy of some initial state. It will be shown that for properly applied fields $\mathbf{S}(\mathbf{x},t)$ and $\mathbf{R}(\mathbf{x},t)$ an arbitrarily large amount of energy can be extracted from some initial state. Therefore even though these fields do not correspond to actual physical objects we believe that the following results are mathematically interesting. The reason we pick these fields is because for particular values of the interaction we obtain an exact solution to the Schrödinger equation. This is demonstrated in the following discussion.

When the interaction is included the Schrödinger equation becomes,

$$i\frac{\partial\left|\Omega(t)\right\rangle}{\partial t}=\hat{H}_T\left|\Omega(t)\right\rangle \tag{2.2}$$

where,

$$\hat{H}_T=\hat{H}+\hat{H}_{int}=\hat{H}-\int\mathbf{S}(\mathbf{x},t)\cdot\hat{\mathbf{A}}(\mathbf{x})d\mathbf{x}-\int\hat{\mathbf{J}}(\mathbf{x})\cdot\mathbf{R}(\mathbf{x},t)d\mathbf{x} \tag{2.3}$$

Now we will solve (2.2) for the following interaction,

$$\mathbf{R}(\mathbf{x},t)=0 \text{ for } t<t_1;\ \mathbf{R}(\mathbf{x},t)=-g(t)\nabla\chi(\mathbf{x}) \text{ for } t_1\le t\le t_2;\ \mathbf{R}(\mathbf{x},t)=0 \text{ for } t>t_2$$

and,

$$\mathbf{S}(\mathbf{x},t)=0 \text{ for } t<t_1;\quad \mathbf{S}(\mathbf{x},t)=\ddot{g}(t)\nabla\chi(\mathbf{x}) \text{ for } t_1\le t\le t_2;\quad \mathbf{S}(\mathbf{x},t)=0 \text{ for } t>t_2$$

where the double dots represent the second derivative with respect to time. In addition to the above $g(t)$ satisfies the following relationship at time $t_2$,

$$\dot{g}(t_2)=0 \text{ and } g(t_2)=-1 \tag{2.4}$$

According to the above expressions the interaction is turned on at time $t_1$ and turned off at time $t>t_2$. During this time energy is exchanged between the quantized fermion-electromagnetic field and the classical fields $\mathbf{S}(\mathbf{x},t)$ and $\mathbf{R}(\mathbf{x},t)$. At some initial time $t_i<t_1$ the state vector is given by $\left|\Omega(t_i)\right\rangle$. We are interested in determining the state vector $\left|\Omega(t_f)\right\rangle$ at some final time $t_f>t_2$. Based on the above remarks the state vector $\left|\Omega(t)\right\rangle$ satisfies,

$$i\frac{\partial|\Omega(t)\rangle}{\partial t}=\hat{H}|\Omega(t)\rangle \text{ for } t<t_1 \tag{2.5}$$

$$i\frac{\partial|\Omega(t)\rangle}{\partial t}=\begin{pmatrix}\hat{H}-\ddot{g}(t)\int\nabla\chi(\mathbf{x})\cdot\hat{\mathbf{A}}(\mathbf{x})d\mathbf{x}\\ +g(t)\int\hat{\mathbf{J}}(\mathbf{x})\cdot\nabla\chi(\mathbf{x})d\mathbf{x}\end{pmatrix}|\Omega(t)\rangle \text{ for } t_1\le t\le t_2 \tag{2.6}$$

$$i\frac{\partial|\Omega(t)\rangle}{\partial t}=\hat{H}|\Omega(t)\rangle \text{ for } t>t_2 \tag{2.7}$$

Since these equations are first order differential equations the boundary conditions at $t_1$ and $t_2$ are,

$$|\Omega(t_1+\varepsilon)\rangle \underset{\varepsilon\to 0}{=} |\Omega(t_1-\varepsilon)\rangle \text{ and } |\Omega(t_2+\varepsilon)\rangle \underset{\varepsilon\to 0}{=} |\Omega(t_2-\varepsilon)\rangle \tag{2.8}$$

The solution to (2.5) is,

$$|\Omega(t)\rangle = e^{-i\hat{H}(t-t_i)}|\Omega(t_i)\rangle \text{ for } t<t_1 \tag{2.9}$$

It is shown in Appendix A that the solution to (2.6) is,

$$|\Omega(t)\rangle = e^{ig(t)\hat{C}}e^{i\dot{g}(t)\hat{D}}e^{iw(t)}e^{-i\hat{H}(t-t_1)}|\Omega(t_1)\rangle \text{ for } t_2\ge t\ge t_1 \tag{2.10}$$

where the operator $\hat{D}$ is defined by,

$$\hat{D}=\int\hat{\mathbf{A}}(\mathbf{x})\cdot\nabla\chi(\mathbf{x})d\mathbf{x} \tag{2.11}$$

and

$$w(t)=\int_{t_1}^{t}\left(\frac{\dot{g}(t')^2}{2}\int|\nabla\chi|^2 d\mathbf{x}+\ddot{g}(t')g(t')\int|\nabla\chi|^2 d\mathbf{x}\right)dt' \tag{2.12}$$

The solution to (2.7) is,

$$|\Omega(t_f)\rangle = e^{-i\hat{H}(t_f-t_2)}|\Omega(t_2)\rangle \text{ where } t_f>t_2 \tag{2.13}$$

Use the boundary conditions (2.8) in the above to obtain,

$$\left|\Omega\left(t_f\right)\right\rangle = e^{-i\hat{H}\left(t_f - t_2\right)} e^{ig\left(t_2\right)\hat{C}} e^{i\dot{g}\left(t_2\right)\hat{D}} e^{iw\left(t_2\right)} e^{-i\hat{H}\left(t_2 - t_i\right)} \left|\Omega\left(t_i\right)\right\rangle \tag{2.14}$$

Use (2.4) in the above to obtain,

$$\left|\Omega\left(t_f\right)\right\rangle = e^{-i\hat{H}\left(t_f - t_2\right)} e^{-i\hat{C}} e^{iw\left(t_2\right)} \left|\Omega_0\left(t_2\right)\right\rangle \tag{2.15}$$

where $\left|\Omega_0\left(t_2\right)\right\rangle$ is defined by,

$$\left|\Omega_0\left(t_2\right)\right\rangle = e^{-i\hat{H}\left(t_2 - t_i\right)} \left|\Omega\left(t_i\right)\right\rangle \tag{2.16}$$

$\left|\Omega_0\left(t_2\right)\right\rangle$ is the state vector that the initial state $\left|\Omega\left(t_i\right)\right\rangle$ would evolve into, by the time $t_2$, in the absence of the interactions. Use (2.14) in (1.12) to show that the energy of the state $\left|\Omega\left(t_f\right)\right\rangle$ is,

$$E\left(\left|\Omega\left(t_f\right)\right\rangle\right) = \left\langle\Omega_0\left(t_2\right)\right| e^{i\hat{C}} \hat{H} e^{-i\hat{C}} \left|\Omega_0\left(t_2\right)\right\rangle \tag{2.17}$$

From the discussion leading up to equation (1.29) we obtain,

$$E\left(\left|\Omega\left(t_f\right)\right\rangle\right) = E\left(\Omega_0\left(t_2\right)\right) - \int \chi\left(\mathbf{x}\right) \nabla \cdot \left\langle\Omega_0\left(t_2\right)\right| \hat{\mathbf{J}}\left(\mathbf{x}\right) \left|\Omega_0\left(t_2\right)\right\rangle d\mathbf{x} \tag{2.18}$$

Now, as before, assume that we select an initial state $\left|\Omega\left(t_i\right)\right\rangle$ so that $\nabla \cdot \left\langle\Omega_0\left(t_2\right)\right| \hat{\mathbf{J}}\left(\mathbf{x}\right) \left|\Omega_0\left(t_2\right)\right\rangle$ is non-zero. Recall that $\left|\Omega_0\left(t_2\right)\right\rangle$ is the state that $\left|\Omega\left(t_i\right)\right\rangle$ evolves into in the absence of interactions. Therefore $E\left(\Omega_0\left(t_2\right)\right) = E\left(\Omega\left(t_i\right)\right)$ and $\left|\Omega_0\left(t_2\right)\right\rangle$ is independent of $\chi\left(\mathbf{x}\right)$. The function $\chi\left(\mathbf{x}\right)$ can take on any value without affecting $\nabla \cdot \left\langle\Omega_0\left(t_2\right)\right| \hat{\mathbf{J}}\left(\mathbf{x}\right) \left|\Omega_0\left(t_2\right)\right\rangle$. Let $\chi\left(\mathbf{x}\right) = \lambda \nabla \cdot \left\langle\Omega_0\left(t_2\right)\right| \hat{\mathbf{J}}\left(\mathbf{x}\right) \left|\Omega_0\left(t_2\right)\right\rangle$ so that (2.18) becomes,

$$E\left(\left|\Omega\left(t_f\right)\right\rangle\right) = E\left(\Omega\left(t_i\right)\right) - \lambda \int \left(\nabla \cdot \left\langle\Omega_0\left(t_2\right)\right| \hat{\mathbf{J}}\left(\mathbf{x}\right) \left|\Omega_0\left(t_2\right)\right\rangle\right)^2 d\mathbf{x} \tag{2.19}$$

Define $\Delta E_{ext}$ as the amount of energy extracted from the quantum state due to its interaction with the classical fields. From the above equation,

$$\Delta E_{ext} = \lambda \int \left( \nabla \cdot \left\langle \Omega_0(t_2) \middle| \hat{\mathbf{J}}(\mathbf{x}) \middle| \Omega_0(t_2) \right\rangle \right)^2 d\mathbf{x} \tag{2.20}$$

Obviously as $\lambda \to \infty$ then $\Delta E_{ext} \to \infty$.

In conclusion, it has been shown that there exist quantum states with less energy than the vacuum state for QED in the temporal gauge. In fact there is no lower bound to the energy of quantum states. If an initial state interacts with properly applied classical fields then it is possible to extract an arbitrarily large amount of energy from the initial state. The classical fields that were applied in this article are mathematical objects that are not assumed to correspond to real physical objects. A possible next step in this research would be to determine if these same results can be obtained through the interaction of real existing fields.

## Appendix A

It will be shown that (2.10) is the solution to (2.6). Take the time derivative of (2.6) and multiply by "i" to obtain,

$$i\frac{\partial}{\partial t}\left|\Omega(t)\right\rangle = -\left(\dot{g}\hat{C} + \dot{w}\right)\left|\Omega(t)\right\rangle - \left|\Omega_a\right\rangle + \left|\Omega_b\right\rangle \tag{A.1}$$

where,

$$\left|\Omega_a\right\rangle = e^{ig\hat{C}}\ddot{g}\hat{D}e^{i\dot{g}\hat{D}}e^{iw(t)}e^{-i\hat{H}(t-t_1)}\left|\Omega(t_1)\right\rangle \tag{A.2}$$

and

$$\left|\Omega_b\right\rangle = e^{ig\hat{C}}e^{i\dot{g}\hat{D}}e^{iw(t)}\hat{H}e^{-i\hat{H}(t-t_1)}\left|\Omega(t_1)\right\rangle \tag{A.3}$$

To evaluate (A.2) we will use the following relationships.

$$e^{ig\hat{C}}\hat{D}=\left(e^{ig\hat{C}}\hat{D}e^{-ig\hat{C}}\right)e^{ig\hat{C}} \tag{A.4}$$

Use (1.6) and (1.25) to obtain,

$$e^{ig\hat{C}}\hat{D}e^{-ig\hat{C}}=\hat{D}+ig\left[\hat{C},\hat{D}\right]=\hat{D}-g\int\left|\nabla\chi\right|^2 d\mathbf{x} \tag{A.5}$$

Use these results in (A.2) to yield,

$$\left|\Omega_a\right\rangle=\ddot{g}\left(\hat{D}-g\int\left|\nabla\chi\right|^2 d\mathbf{x}\right)\left|\Omega(t)\right\rangle \tag{A.6}$$

Next evaluate (A.3). Use (1.25) and the commutation relationships to obtain,

$$e^{i\dot{g}\hat{D}}\hat{H}e^{-i\dot{g}\hat{D}}=\hat{H}+i\dot{g}\left[\hat{D},\hat{H}\right]-\frac{\dot{g}^2}{2}\left[\hat{D},\left[\hat{D},\hat{H}\right]\right] \tag{A.7}$$

where,

$$\left[\hat{D},\hat{H}\right]=\int\left[\hat{\mathbf{A}},\hat{H}\right]\cdot\nabla\chi d\mathbf{x}=\int\left[\hat{\mathbf{A}},\hat{H}_{0,M}\right]\cdot\nabla\chi d\mathbf{x}=-i\hat{C} \tag{A.8}$$

and,

$$\left[\hat{D},\left[\hat{D},\hat{H}\right]\right]=-i\left[\hat{D},\hat{C}\right]=-i\left[\int\hat{\mathbf{A}}\cdot\nabla\chi d\mathbf{x},\int\hat{\mathbf{E}}\cdot\nabla\chi d\mathbf{x}\right]=-\int\left|\nabla\chi\right|^2 d\mathbf{x} \tag{A.9}$$

Therefore,

$$e^{i\dot{g}\hat{D}}\hat{H}e^{-i\dot{g}\hat{D}}=\hat{H}+\dot{g}\hat{C}+\frac{\dot{g}^2}{2}\int\left|\nabla\chi\right|^2 d\mathbf{x} \tag{A.10}$$

Use this in (A.3) to obtain,

$$\left|\Omega_b\right\rangle=e^{ig\hat{C}}\left(\hat{H}+\dot{g}\hat{C}+\frac{\dot{g}^2}{2}\int\left|\nabla\chi\right|^2 d\mathbf{x}\right)e^{i\dot{g}\hat{D}}e^{iw(t)}e^{-i\hat{H}(t-t_1)}\left|\Omega(t_1)\right\rangle \tag{A.11}$$

To evaluate this further use,

$$e^{ig\hat{C}}\hat{H}e^{-ig\hat{C}}=\hat{H}+ig\left[\hat{C},\hat{H}\right]=\hat{H}+g\left(\int\hat{\mathbf{J}}\cdot\nabla\chi d\mathbf{x}\right) \tag{A.12}$$

where we have used,

$$\left[\hat{C},\hat{H}\right]=-i\int\hat{\mathbf{J}}\cdot\nabla\chi d\mathbf{x} \tag{A.13}$$

and the fact that $\left[\hat{C},\left[\hat{C},\hat{H}\right]\right]=0$. Therefore,

$$\left|\Omega_b\right\rangle=\left(\hat{H}+g\left(\int\hat{\mathbf{J}}\cdot\nabla\chi d\mathbf{x}\right)+\dot{g}\hat{C}+\frac{\dot{g}^2}{2}\int\left|\nabla\chi\right|^2 d\mathbf{x}\right)\left|\Omega(t)\right\rangle \tag{A.14}$$

Use this along with (A.11) and (A.6) in (A.1) to obtain,

$$i\frac{\partial}{\partial t}\left|\Omega(t)\right\rangle=\left\{\begin{array}{l}\left(\hat{H}+g\left(\int\hat{\mathbf{J}}\cdot\nabla\chi d\mathbf{x}\right)+\dot{g}\hat{C}+\frac{\dot{g}^2}{2}\int\left|\nabla\chi\right|^2 d\mathbf{x}\right)-\left(\dot{g}\hat{C}+\dot{w}\right)\\ -\ddot{g}\left(\hat{D}-g\int\left|\nabla\chi\right|^2 d\mathbf{x}\right)\end{array}\right\}\left|\Omega(t)\right\rangle \tag{A.15}$$

Rearrange terms and do some simple algebra to obtain,

$$i\frac{\partial}{\partial t}\left|\Omega(t)\right\rangle=\left\{\begin{array}{l}\left(\hat{H}+g\int\hat{\mathbf{J}}\cdot\nabla\chi d\mathbf{x}-\ddot{g}\int\hat{\mathbf{A}}\cdot\nabla\chi d\mathbf{x}\right)\\ -\left(\dot{w}-\frac{\dot{g}^2}{2}\int\left|\nabla\chi\right|^2 d\mathbf{x}-\ddot{g}g\int\left|\nabla\chi\right|^2 d\mathbf{x}\right)\end{array}\right\}\left|\Omega(t)\right\rangle \tag{A.16}$$

Now let

$$\dot{w}=\frac{\dot{g}^2}{2}\int\left|\nabla\chi\right|^2 d\mathbf{x}+\ddot{g}g\int\left|\nabla\chi\right|^2 d\mathbf{x} \tag{A.17}$$

to obtain

$$i\frac{\partial}{\partial t}\left|\Omega(t)\right\rangle=\left(\hat{H}+g\int\hat{\mathbf{J}}\cdot\nabla\chi d\mathbf{x}-\ddot{g}\int\hat{\mathbf{A}}\cdot\nabla\chi d\mathbf{x}\right)\left|\Omega(t)\right\rangle \tag{A.18}$$

which is (2.6) in the text. This completes the proof.